\documentclass[12pt]{article}
\usepackage{amsmath}
\usepackage{amssymb}
\usepackage[mathscr]{eucal}
\usepackage{authblk}
\usepackage[nocompress]{cite}
\usepackage{enumitem}
\usepackage{pdfsync}
\usepackage[
bookmarks=true,
backref=true,
colorlinks=true,
linkcolor=red,
citecolor=blue,
urlcolor=green]{hyperref}

\newcommand{\gen}[1]{\partial_{#1}}

\newcommand{\curl}[1]{ \left\{#1\right\} }
\newcommand*{\email}[1]{\normalsize\href{mailto:#1}{#1}\par}

\newtheorem{example}{Example}[section]

\newtheorem{thm}{Theorem}
\newtheorem{cor}{Corollary}
\newtheorem{rmk}{Remark}

\numberwithin{equation}{section}
\numberwithin{thm}{section}
\numberwithin{lemma}{section}
\numberwithin{prop}{section}
\numberwithin{cor}{section}
\numberwithin{rmk}{section}
\numberwithin{defn}{section}
\DeclareMathOperator{\Sl}{sl}
\DeclareMathOperator{\sech}{sech}

\DeclareMathOperator{\SL}{SL}

\DeclareMathOperator{\So}{so}

\DeclareMathOperator{\erf}{erf}

\DeclareMathOperator{\ort}{o}
\DeclareMathOperator{\Heis}{H}
\newcommand{\lie}{\mathfrak g}
\newcommand{\nil}{\mathfrak {h}}

\newcommand{\D}{\mathrm D}

\begin{document}

\title{A general class of  invariant diffusion processes in one dimension}

\author{F.~G\"ung\"or }

\affil{Department of Mathematics, Faculty of Science and Letters, Istanbul Technical University, 34469 Istanbul, Turkey\\  \email{gungorf@itu.edu.tr} }

\date{}

\maketitle

\abstract{This paper extends a previously established semi-invariant criterion ensuring equivalence by point transformations to Lie's canonical forms and  presence of  non-trivial Lie point symmetries (at least four-dimensional)  for diffusion processes  in the case of spatial dependence to those with the added temporal dependence in the coefficients.   We apply this criterion  to various physically important  models arising in   applications and demonstrate how our approach allows determination of non-trivial symmetries,  corresponding symmetry groups and interesting exact solutions.
}

\bigskip
Keywords: Diffusion process, Brownian  motion, Lie symmetry group, transformation formula to canonical forms, gauge invariance, fundamental solution

\section{Introduction}

A class of partial differential equations
\begin{equation}\label{mainLPE}
u_t=a(x)u_{xx}+b(x)u_x+c(x)u, \quad a> 0,\quad x\in \mathbb{R}, \quad t>0,
\end{equation}
was studied in an earlier paper  \cite{Guengoer2018} from the point of view of local equivalence and invariance.
Here $a$, $b$, $c$ are arbitrary smooth functions of $x$. The coefficients $a$ and $b$ are called diffusion and drift functions. Partial differential equations (PDEs) from the class \eqref{mainLPE} are basic  modelling equations for many applications in applied mathematics and science ranging from  stochastic processes (Brownian motion),  applied probability theory, financial mathematics to population genetics.

The purpose of this paper is  to  concentrate on the most general case of the coefficients depending not only on space but also on time variable
\begin{equation}\label{generalpde}
 u_t=a(x,t)u_{xx}+b(x,t)u_x+c(x,t)u,  \quad a>0.
\end{equation}
Below we introduce a new (nonlocal) invariant of the coefficients $a,b,c$ which will play a central role in characterising PDEs equivalent to  standard heat equation.

We observe that any PDE of the form \eqref{mainLPE} admits the time translation symmetry $u(x,t)\to u(x,t+t_0)$ and the scaling symmetry $u(x,t)\to \lambda u(x,t)$ for any constant $\lambda$.  This class has as well the linear superposition principle. These symmetries are the trivial ones.  In this paper we shall be interested only in  non-trivial symmetries (symmetries other than trivial ones).

The symmetry group of the standard heat equation where $a=1, b=c=0$ was first computed by Lie \cite{Lie1881} and later recomputed by Appell \cite{Appell1982}. A complete symmetry group classification of the superclass of \eqref{mainLPE} where the coefficients depend on $x$ and $t$  was also performed by Lie himself in \cite{Lie1881}. The historical details have been discussed in \cite{Olver1991}. For the most general class there are two canonical forms which admit non-trivial symmetry groups
\begin{equation}\label{canon}
  u_t=u_{xx},  \quad u_t=u_{xx}+\frac{\mu}{x^2}u,  \quad \mu\ne 0.
\end{equation}
Alternative derivations of this classification  can be found in Refs. \cite{Ovsiannikov1982} or \cite{Guengoer2018}.
The first canonical form (the heat equation) has the Lie group structure
$G=\SL(2,\mathbb{R})\ltimes \Heis(1)$, while the second canonical one has  the Lie group structure $G=\SL(2,\mathbb{R})\ltimes \mathbb{R}$, where $\ltimes$ denotes a semi-direct product, $\Heis(1)$ is the 3-dimensional Heisenberg group.

The structure of this paper is as follows. In Section 2, we review the Lie point symmetry analysis of \eqref{mainLPE}. Solutions of the invariance condition as a Riccati equation in the independent variable $I(x)$ has been discussed. Section 3 is devoted to five examples of PDEs in connection with the study of Brownian motion and mathematical finance. The results of Section 2 are applied to them to construct solutions of physical interest.

\section{Equivalence transformations and Lie point  symmetries}\label{point-symm}

\subsection{An invariant criterion of transformability to the heat equation}
The group of equivalence transformations of the general PDE \eqref{generalpde}
is known to be given by invertible point transformations of the form
\begin{equation}\label{equiv-trans}
  \tilde{t}=T(t),  \quad \tilde{x}=X(x,t),  \quad \tilde{u}=\theta(x,t)u,
\end{equation}
where $\dot{T}X_x\ne 0$  and $\theta(x,t)\neq 0$.

In a previous work \cite{Guengoer2018} where the coefficients $a,b,c$ are only spatially varying in the parabolic equation \eqref{mainLPE} we introduced the following two quantities in terms of the coefficients
\begin{equation}\label{IJ}
I(x)=\int^{x}\frac{ds}{\sqrt{a(s)}},  \quad J(x)=\frac{1}{\sqrt{a}}\left[\frac{a'(x)}{2}-b(x)\right]=\frac{d}{dx}(\sqrt{a(x)})-\frac{b(x)}{\sqrt{a(x)}}
\end{equation}
Another function that  played a key role in the identification of point symmetries and mapping of \eqref{mainLPE} to canonical forms was
\begin{equation}\label{semi-inv}
  K(x):= F(a,b,c,a_x,a_{xx},b_x)=\frac{1}{2}\sqrt{a}J'(x)-\frac{1}{4}J^2(x)+c(x).
\end{equation}
This function $K(x)$ is a (second order) semi-invariant (See Subsection \ref{sect-semi}) of  \eqref{mainLPE}  as it remains unaltered under the rescaling of the dependent variable: $\tilde{u}=\theta(x,t)u$, where the gauge factor $\theta\ne 0$ ($\theta_t=0$) is arbitrary.

When we allow the coefficients to depend on both temporal and spatial variables
\begin{equation}\label{coeff}
  a=a(x,t), \quad b=b(x,t),  \quad c=c(x,t),
\end{equation}
we need to incorporate additional terms involving time derivatives of $a$ and $b$ into   $K(x)$.
We introduce the following functions
\begin{equation}\label{eq:IJRB}
\begin{aligned}
I(x,t)&=\int_{x_0}^{x}\frac{ds}{\sqrt{a(s,t)}},&
J(x,t)&=\frac{\partial}{\partial x}\sqrt{a(x,t)}-\frac{b(x,t)}{\sqrt{a(x,t)}},\\
R(x,t)&=\int_{x_0}^{x}\frac{I_t(s,t)}{\sqrt{a(s,t)}}\,ds, & B(x,t)&=\int_{x_0}^{x}\frac{b(s,t)}{a(s,t)}\,ds.
\end{aligned}
\end{equation}
Unless otherwise specified, partial derivatives with respect to $t$ of functions of $(x,t)$ are taken at fixed $x$.

Also, we replace $K(x)$ by $K(x,t)$
\begin{multline}\label{semi-inv-gen}
   K(x,t):= \frac{1}{2}\sqrt{a}J_x(x,t)-\frac{1}{4}J^2(x,t)+c(x,t) \\
  -\frac{a_t(x,t)}{2a(x,t)}
+\frac14I_t(x,t)^2+\frac{1}{2}(B_t(x,t)-R_t(x,t)).
\end{multline}
If, in particular, $a(x,t)=A(t)a(x)$ then by a reparametrization of $t$ we can set $A(t)=1$. So
we have $a_t=0$, which implies $I_t=R=0$. Thus, $K(x,t)$ simplifies to
\begin{equation}\label{K-reduced}
   K(x,t)=\frac{1}{2}\sqrt{a}J_x(x,t)-\frac{1}{4}J^2(x,t)+c(x,t)+\frac{1}{2}\int^{x}\frac{b_t(s,t)}{a(s)}ds.
\end{equation}
When $b_t=0$ the integral vanishes, recovering the expression in \eqref{semi-inv}.



We proceed along the similar lines of \cite{Guengoer2018} and use the transformation \eqref{equiv-trans} to normalize $a(x,t)$ and transform away the coefficient $b(x,t)$ and obtain the following result.

\begin{thm}\label{thm-2.2}
An equation from the class \eqref{generalpde} can be transformed to the heat equation if and only if the semi-invariant $K$ equals a quadratic polynomial in $I$
\begin{equation}\label{non-trivial-sym-cond}
K(x,t)=c_2(t)I^2(x,t)+c_1(t)I(x,t)+c_0(t),
\end{equation}
where $c_2$, $c_1$ and $c_0$ are certain functions of $t$. Furthermore,
the transformation mapping \eqref{mainLPE} to the standard heat equation $\tilde{u}_{\tilde{t}}=\tilde{u}_{\tilde{x}\tilde{x}}$ is given by
\begin{equation}\label{ToHeatTr}
\tilde{t}=T(t),  \quad \tilde{x}=\sqrt{\dot{T}}(I+\omega(t)), \quad \dot{T}>0,
\end{equation}
\begin{equation}\label{ToHeatTr-u}
\begin{aligned}
u(x,t)={}&C[\dot T(t)a(x,t)]^{1/4}\exp\Bigg[
-\frac12(B(x,t)-R(x,t))+\frac{\ddot T}{8\dot T}(I+\omega)^2
+\frac{\dot\omega}{2}I\\
&+\int^t\left(c_0(s)+\frac{\dot\omega(s)^2}{4}
-c_2(s)\omega(s)^2\right)ds\Bigg]
\tilde u(\tilde x,\tilde t),
\end{aligned}
\end{equation}
where $C$ is an arbitrary nonzero constant and $T$, $\omega$ are  solutions to the set of ODEs
\begin{equation}\label{ODE-pair}
  \curl{T,t}=8c_2,  \quad \ddot{\omega}+4c_2\omega=2 c_1,
\end{equation}
where $\curl{T,t}$ denotes the Schwarzian derivative of $T$. This transformation in general depends on six arbitrary constants (See Example \ref{CIR-ex} for an application).

We comment that the same transformation can be used to map to the second canonical form \eqref{canon} of the heat equation which in this case we must have $c_1=0$ and $\omega=0$.

\end{thm}

We recall that the ratio $\varrho=\ddot{T}/\dot{T}$ satisfies the Riccati equation
\begin{equation}\label{Riccati}
  \dot{\varrho}-\frac{1}{2}\varrho^2=8c_2.
\end{equation}
There are three possible cases for the solution of the Schwarzian equation $\curl{T,t}=8c_2$ depending on the sign of $c_2$:
\begin{enumerate}
  \item $c_2=0$:
  $$T(t)=M(t)=\frac{\alpha t+\beta}{\gamma t+\delta}, \quad \Delta=\alpha \delta-\beta\gamma> 0, \quad \varrho=\frac{\ddot{T}}{\dot{T}}=\frac{-2\gamma}{\gamma  t+\delta}.$$
   \item $c_2=-\lambda^2$, $\lambda>0$:
   $$T(t)=\frac{\alpha \cosh 2\lambda t+\beta \sinh 2\lambda t}{\gamma \cosh 2\lambda t+\delta \sinh 2\lambda t},  \quad \dot{T}=-2\Delta \lambda(\gamma \cosh 2\lambda t+\delta \sinh 2\lambda t)^{-2}.$$
     $$ \varrho=\frac{\ddot{T}}{\dot{T}}=-4\lambda \tanh(2\lambda t+\varepsilon), \quad \tanh \varepsilon=\frac{\delta}{\gamma}, \quad \Delta<0.$$
  \item $c_2=\lambda^2$, $\lambda>0$:
  $$T(t)=\frac{\alpha \cos 2\lambda t+\beta \sin 2\lambda t}{\gamma \cos 2\lambda t+\delta \sin 2\lambda t},  \quad \dot{T}=-2\Delta \lambda(\gamma \cos 2\lambda t+\delta \sin 2\lambda t)^{-2}.$$
     $$ \varrho=\frac{\ddot{T}}{\dot{T}}=4\lambda \tan(2\lambda t+\varepsilon), \quad \tan \varepsilon=\frac{\delta}{\gamma}, \quad \Delta<0.$$
\end{enumerate}

We point out that as $\alpha, \beta, \gamma, \delta$ are defined up to nonzero constant multiplicative and $\dot{T}>0$, we can always put $\Delta=\alpha \delta-\beta \gamma=\pm 1$.

Formula \eqref{ToHeatTr-u} means that if $u(x,t)$ satisfies \eqref{mainLPE}, then $\tilde{u}(\tilde{x},\tilde{t})$ will satisfy the heat equation $\tilde{u}_{\tilde{t}}=\tilde{u}_{\tilde{x}\tilde{x}}$. In some cases, this transformation can map the trivial solutions (a constant one) to interesting solutions like fundamental solutions (see Section \ref{app}).

If, in particular,  we require $u$ to be solution of the heat equation, then Theorem \ref{thm-2.2} provides us with the symmetry group of this PDE (see Theorem 2 of \cite{KovalPopovych2023}  for an alternative derivation of this result). It is sufficient to choose
$a=1$, $b=c=0$, which implies $K=0$ and $c_2=c_1=c_0$. So we have to solve $\curl{T,t}=0$ and $\ddot{\omega}=0$. $T(t)$ are Möbius transformations of $t$ and $\omega(t)=vt+x_0$.

\begin{cor}
  The complete symmetry group of the heat equation, up to adding an arbitrary solution to $u$,  is given by
\begin{equation}\label{symmgr-canon1}
\begin{split}
&\tilde{t}=\frac{\alpha t+\beta}{\gamma t+\delta},  \quad \tilde{x}=\frac{x+vt+x_0}{\gamma t+\delta}, \\
&\tilde{u}=C_0 |\gamma t+\delta|^{1/2}\exp\left[\frac{\gamma(x+vt+x_0)^2}{4(\gamma t+\delta)}-\frac{v  }{2}x-\frac{v^2}{4}t\right]u(\tilde{x},\tilde{t}),
\end{split}
\end{equation}
where $C_0$,$v, x_0$ and $\alpha, \beta, \gamma, \delta$ such that $\alpha \delta-\beta \gamma=1$ are arbitrary parameters.
\end{cor}

We note that in the case of  four-dimensional symmetry algebra,  $\omega=0$ ($c_1=0$). So we only have to solve $\curl{T,t}=8c_2$. Now we can use formula \eqref{ToHeatTr} to obtain a point transformation mapping to the second canonical form (heat equation with inverse square potential). This gives us the symmetry group of this PDE for the choice
\begin{equation}\label{inv-square-tr}
  a=1,  \quad b=0, \quad  c=\frac{\mu}{x^2}.
\end{equation}
Then,  $J=0$, $K=c$ and form the condition \eqref{L4}, it follows that $c_2=0$ ($c_1$ is already zero).

\begin{cor}
  The complete symmetry group of the second canonical form consists of point transformations of the form
\begin{equation}\label{symmgr-canon2}
\begin{split}
&\tilde{t}=\frac{\alpha t+\beta}{\gamma t+\delta},  \quad \tilde{x}=\frac{x}{\gamma t+\delta}, \\
&\tilde{u}=C_0 |\gamma t+\delta|^{1/2}\exp\left[\frac{\gamma x^2}{4(\gamma t+\delta)}\right]u(\tilde{x},\tilde{t}),
\end{split}
\end{equation}
where $C_0$ and $\alpha, \beta, \gamma, \delta$ such that $\alpha \delta-\beta \gamma=1$ are arbitrary parameters.
\end{cor}
See Theorem 9 of \cite{KovalBihloPopovych2023} for an alternative construction of \eqref{symmgr-canon2}.

\subsection{Semi-invariants}\label{sect-semi}

By an invariant of \eqref{generalpde} we shall mean a differential function $F(C,C^{(s)})$ of the coefficients $C=(a,b,c)$ and their space-time derivatives $C^{(s)}$ up to some order $s$, which is invariant under the change of dependent and independent variables \eqref{equiv-trans} so that $F(\tilde{C},\tilde{C}^{(s)})=F(C,C^{(s)})$, where $\widetilde{C}=(\tilde{a},\tilde{b},\tilde{c})$ denotes the transformed coefficients.

If   only invariance  under the rescaling of the dependent variable $u$ by a nonvanishing function $\theta(x,t)=e^{f(x,t)}$ (the so-called gauge factor) exists, then $F$ will be called a semi-invariant of \eqref{generalpde}. Such a transformation changes the coefficients of the PDE \eqref{generalpde} to
\begin{equation}\label{new-coeff}
  \tilde{a}=a,  \quad \tilde{b}=b-2a f_x,  \quad \tilde{c}=c+f_t-bf_x+a(f_x^2-f_{xx}).
\end{equation}
The first relation implies that $a$ is a semi-invariant.

Let us write $J=(\sqrt a)_x-b/\sqrt a$ and $B=\int_{x_0}^x (b/a)(s,t)\,ds$ and look at the gauge transformation of $K(x,t)$. Then
\[
 \widetilde J=J+2\sqrt a\,f_x,\qquad
 \widetilde B=B-2f(x,t)+2f(x_0,t).
\]
The functions $I$ and $R$ in \eqref{eq:IJRB} depend only on $a$ and are gauge invariant. Setting $S=\tfrac12\sqrt a\,J_x-\tfrac14J^2+c$, direct substitution gives $\widetilde S=S+f_t(x,t)$. In \eqref{eq:IJRB} the contribution from $\tfrac12\widetilde B_t$ cancels the term at $x$ and retains the term at $x_0$. Thus for general $a=a(x,t)$,
\[
 \widetilde K(x,t)=K(x,t)+f_t(x_0,t),\qquad
 \widetilde K_x=K_x.
\]
The quadratic criterion \eqref{non-trivial-sym-cond} is preserved since the shift in the gauge transformation of $K$ is absorbed into $c_0(t)$.

Now let us assume $a_t=0$. Putting $P=b/(2a)$ and expanding $J=(\sqrt a)_x-2\sqrt a\,P$ gives
\begin{equation}\label{S}
   S=Q(x,t)+\frac{a_{xx}}4-\frac{3a_x^2}{16a},  \qquad Q=c-aP_x-aP^2.
\end{equation}
Consequently, when $a_t=0$,
\begin{equation}\label{Kx}
 K_x=H+\D_x\!\left(\frac{a_{xx}}4-\frac{3a_x^2}{16a}\right),\qquad
 H=\D_x(c-aP_x-aP^2)+P_t.
\end{equation}

The same formula for $H$ gives a local semi-invariant without the restriction $a_t=0$. 
Equation \eqref{new-coeff} gives $\widetilde P=P-f_x$ and $\widetilde Q=Q+f_t$. Equality of the mixed derivatives of $f$ therefore yields
$$
H(x,t)=Q_x+P_t,\qquad \widetilde H=H.
$$
For two coefficient triples $C=(a,b,c)$ and $\widetilde{C}=(\tilde{a},\tilde{b},\tilde{c})$ with the same $a$, the equations for a gauge relating them are
$$
 f_x=P-\widetilde P,\qquad f_t=\widetilde Q-Q.
$$
Their compatibility condition is precisely $\widetilde H=H$. Two members of the class \eqref{generalpde} can be mapped into each other if and only if the (second-order) semi-invariant
$H(x,t)$
has the same value. On a sufficiently small simply connected neighbourhood this condition gives a smooth $f$, unique up to an additive constant.

An infinitesimal approach can be used to verify that the set   $\{a,H\}$ actually forms a basis of  second order semi-invariants up to $a$ and its derivatives when $a_t=0$. Since the gauge fixes both independent variables, total differentiation with respect to either $x$ or $t$ preserves semi-invariance. Arbitrary function of this set is again a semi-invariant and any  higher order ones can be generated by differentiating this basis with respect to time and space coordinates.

In this approach, we look at the infinitesimal changes in the coefficients $a,b,c$ of the original PDE under the infinitesimal transformations corresponding to   the rescaling of $u$
\begin{equation}\label{inf-u}
  \tilde{x}=x, \quad \tilde{t}=t,  \quad \tilde{u}=e^{\varepsilon \eta(x,t)}u=u+\varepsilon \eta(x,t)u+O(\varepsilon^2).
\end{equation}
Putting $f=\varepsilon \eta$ in \eqref{new-coeff} and keeping terms linear in $\varepsilon$ we see that the coefficients transform infinitesimally as follows
\begin{equation}\label{tr-coeff}
  \tilde{a}=a,  \quad \tilde{b}=b-2a\varepsilon \eta_x+O(\varepsilon^2),  \quad \tilde{c}=c+\varepsilon(\eta_t-a\eta_{xx}-b\eta_x)+O(\varepsilon^2).
\end{equation}
Here $\eta$ is an arbitrary smooth function. The gauge multiplier $e^{\varepsilon\eta}$ is nonzero even at points where $\eta$ vanishes.
The associated infinitesimal generator of \eqref{tr-coeff} in the space of coefficients is represented by
\begin{equation}\label{inf-gen-coeff}
  X=-2a\eta_x\gen b+(\eta_t-a\eta_{xx}-b\eta_x)\gen c.
\end{equation}
The action of second prolongation  of $X$ on $K_x$ of \eqref{Kx} vanishes for any choice of $\eta(x,t)\ne 0$ everywhere on the second jet space of the functions $a,b,c$. In other words, the semi-invariant $K_x$ is a differential invariant of the infinitesimal transformation \eqref{tr-coeff}. Any other semi-invariant is a function of $a, K_x$ and their derivatives with respect to $x$. For $a=a(x)$, the differential order of $K_x$ is generally three, although it has a representative of order two modulo derivatives of $a$.

The term $a_{xxx}/4$ in \eqref{Kx} makes $K_x$ a third-order function on the full coefficient jet space. The second prolongation annihilates $H$. The third prolongation annihilates $K_x$, since the additional term depends only on $a$ and its derivatives.

Second order semi-invariants of the  PDE \eqref{mainLPE} with  general coefficients \eqref{coeff} using an infinitesimal approach was calculated in \cite{Ibragimov2002}.  Later, in Ref. \cite{JohnpillaiMahomed2001}  a fifth order singular invariant equation of the PDE invariant under the full equivalence group in terms of the second order invariant $H$ and the zeroth order invariant $a$ along with its higher derivatives was constructed with the same method. The authors of  \cite{JohnpillaiMahomed2001} demonstrated that a necessary and sufficient condition for the PDE to be transformable to the heat equation is that a fifth order singular invariant equation be satisfied.

Below it will be clear that the use of our  semi-invariant  \eqref{semi-inv-gen}  calculated through a direct approach  rather  than far too complicated singular invariant equation of the full equivalence group calculated in \cite{JohnpillaiMahomed2001} is much simpler for checking a priori reducibility to the first and second canonical forms.    On the other hand,  the singular invariant equation of \cite{JohnpillaiMahomed2001}  will  impose conditions as fifth, fourth and third  order  nonlinear ODEs, respectively when we let one of the coefficients $a,b,c$ remain free. We refer to Example \ref{volat} to see how our criterion  in comparison to the singular invariant condition simply reveals the required conditions for the PDE under study to possess heat or Schrödinger algebras as Lie point symmetry algebras.

\subsection{Symmetry algebra}
Now we can turn to discuss the infinitesimal symmetries of \eqref{mainLPE}. For completeness, we shall recall briefly some relevant results of \cite{Guengoer2018}. We take this opportunity to correct a number of  misprints and an omission therein.

A general element of the symmetry algebra of \eqref{mainLPE} will be a  vector field of the form
\begin{equation}\label{VF}
v=\tau(t)\gen t+\xi(x,t)\gen x+\phi(x,t)u\gen u,
\end{equation}
where the coefficients $\tau, \xi, \phi$ are determined by applying Lie's infinitesimal invariance algorithm \cite{Guengoer2018} (see \cite{Olver1991} for the details of the general theory). The coefficients $\xi, \phi$ of $v$ are given by
\begin{equation}\label{xi}
  \xi(x,t)=\sqrt{a}\left(\frac{1}{2}\dot{\tau}I(x)+\rho(t)\right),
\end{equation}
and
\begin{equation}\label{phi}
  \phi(x,t)=-\frac{1}{8}\ddot{\tau}I^2-\frac{1}{2}\dot{\rho}I+\frac{1}{4}\dot{\tau} IJ+\frac{1}{2}\rho J+\sigma(t).
\end{equation}
The undetermined functions $\tau, \rho, \sigma$ must satisfy the classifying (determining) ODE
\begin{equation}\label{comp}
\frac{1}{8}\dddot{\tau}I^2-\frac{1}{4}\ddot{\tau}+\dot{\tau}\left(K(x)+\frac{1}{2}\sqrt{a}K'(x)I(x)\right)+\frac{1}{2}\ddot{\rho} I + \rho \sqrt{a} K'(x)  -\dot{\sigma}=0.
\end{equation}
Analysis of \eqref{comp} reveals two possibilities for the invariant $K$ depending on $\rho$:
\begin{align}\label{K}
1)\quad   & \rho=0:\quad K(x)=f_4(I)=\mu I^{-2}+c_2 I^2+c_0,\quad \mu\ne 0,\\
2)\quad   & \rho\not=0:\quad K(x)=f_6(I)=c_2 I^2+c_1 I+c_0 \label{K-2}.
\end{align}

Substituting the above forms of $K$ into \eqref{comp} and splitting with respect to the powers of $I$ result in the system of ODEs for functions $\tau$, $\rho$, and $\sigma$
\begin{equation}\label{deteqs}
\dddot{\tau}+16c_2\dot{\tau}=0,  \quad \ddot{\rho}+4c_2 \rho=-3 c_1\dot{\tau},  \quad \dot{\sigma}=-\frac{1}{4}\ddot{\tau}+c_0\dot{\tau}+c_1 \rho.
\end{equation}

\subsection{Four-dimensional symmetry algebra}\label{Subsection4-dim}
In the  case $\rho=0$,
the general solution of this system will depend on 4 arbitrary independent constants which lead to the following bases
for the corresponding algebras.

Depending on the sign of $c_2$ we have three possible cases for the infinitesimal symmetries. The symmetry vector fields in some convenient basis together with the commutation relations among them are given as follows:
\begin{enumerate}
\item $c_2=0$.
\begin{equation}\label{4-dim-c20}
\begin{split}
&v_1=\gen t,\\
&v_2=t\gen t+\frac{1}{2}\sqrt{a}I\gen x+(c_0t+\frac{1}{4}IJ)u\gen u,\\
&v_3=t^2\gen t+t\sqrt{a}I\gen x+\frac{1}{4}[2(2c_0t-1)t-I^2+2tIJ]u\gen u,\\
&v_4=u\gen u.
\end{split}
\end{equation}
The non-zero commutation relations are
\begin{equation}\label{comm-4-dim-c20}
\begin{split}
&[v_1, v_2]=v_1+c_0v_4,\\
&[v_1, v_3]=2v_2-\frac{1}{2}v_4, \quad [v_2, v_3]=v_3.
\end{split}
\end{equation}

\item $c_2=-\kappa^2$, $\kappa>0$.
\begin{equation}\label{4-dim-c2n}
\begin{split}
&v_1=\gen t,\\
&v_2=e^{4\kappa t}\left[\gen t+2\kappa \sqrt{a}I\gen x-(-c_0+\kappa+2\kappa^2I^2-\kappa IJ)u\gen u\right],\\
&v_3=e^{-4\kappa t}\left[\gen t-2\kappa \sqrt{a}I\gen x+(c_0+\kappa-2\kappa^2I^2-\kappa IJ)u\gen u\right],\\
&v_4=u\gen u.
\end{split}
\end{equation}
The non-zero commutation relations are
\begin{equation}\label{comm-4-dim-c2n}
\begin{split}
&[v_1, v_2]=4\kappa v_2,\\
&[v_1, v_3]=-4\kappa v_3, \quad [v_2, v_3]=-8\kappa v_1-8c_0 \kappa v_4.
\end{split}
\end{equation}
The basis elements $v_2$ and $v_3$ of \eqref{4-dim-c2n} can be expressed in terms of hyperbolic functions $\{\cosh 4\kappa t, \sinh 4\kappa t\}$ by taking the combination
$$v_2\to \frac{1}{2}(v_2+v_3),  \quad v_3\to \frac{1}{2}(v_2-v_3).$$

\item $c_2=\kappa^2$, $\kappa>0$.
\begin{equation}\label{4-dim-c2p}
\begin{split}
&v_1=\gen t,\\
&v_2=\cos 4\kappa t\gen t-2\kappa \sin 4\kappa t \sqrt{a}I\gen x+[ \cos 4\kappa t(c_0+2 \kappa^2 I^2)\\
& \quad +\kappa \sin 4\kappa t(1- IJ)]u\gen u,\\
&v_3=\sin 4\kappa t\gen t+2\kappa \cos 4\kappa t \sqrt{a}I\gen x+[ \sin 4\kappa t(c_0+2 \kappa^2 I^2)\\
& \quad -\kappa \cos 4\kappa t(1- IJ)]u\gen u,\\
&v_4=u\gen u.
\end{split}
\end{equation}
The non-zero commutation relations are
\begin{equation}\label{comm-4-dim-c2p}
\begin{split}
&[v_1, v_2]=-4\kappa v_3,\\
&[v_1, v_3]=4\kappa v_2, \quad [v_2, v_3]=4\kappa (v_1+c_0 v_4).
\end{split}
\end{equation}

\end{enumerate}

\begin{rmk}\label{rmk-lie4}
After suitable central shifts of the basis elements, the vector fields generate a three-dimensional subalgebra isomorphic to $\Sl(2,\mathbb R)$.
For example, this is the case for the algebra with basis \eqref{comm-4-dim-c20},  which is easily seen by the change of basis
\[
e_1=v_1+c_0v_4,\qquad e_2=2v_2-\frac12v_4,\qquad e_3=v_3.
\]
These generators satisfy
\[
[e_1,e_2]=2e_1,\qquad [e_1,e_3]=e_2,\qquad [e_2,e_3]=2e_3.
\]
The entire algebra is isomorphic to the direct sum $\lie_4=\Sl(2,\mathbb{R})\oplus \mathbb R$
The same isomorphism is also true for the algebras spanned by \eqref{4-dim-c2n} and \eqref{4-dim-c2p}.
\end{rmk}

\subsection{Six-dimensional symmetry algebra}\label{Subsection6-dim}

In the  case $\rho\not=0$, the general solution of the system \eqref{deteqs}
will depend on 6 arbitrary independent constants which lead to the following bases
for the corresponding algebras.

\begin{enumerate}
\item $c_2=0$.
\begin{equation}\label{6-dim-c20}
\begin{split}
&v_1=\gen t,\\
&v_2=t\gen t+\frac{1}{2}\sqrt{a}(I-3c_1t^2)\gen x+\frac{1}{4}[I(6c_1 t+J)+t(4c_0-2c_1^2t^2-3c_1tJ)]u\gen u,\\
&v_3=t^2\gen t+t\sqrt{a}(I-c_1 t^2)\gen x-\frac{1}{4}[I^2-2tI(3c_1 t+J)+t(2-4c_0 t+c_1^2 t^3+2c_1 t^2 J)]u\gen u,\\
&v_4=t\sqrt{a}\gen x+\frac{1}{2}[-I+t(c_1 t+J)]u\gen u,\\
&v_5=\sqrt{a}\gen x+\frac{1}{2}[2c_1t+J]u\gen u,\\
&v_6=u\gen u.
\end{split}
\end{equation}
The non-zero commutation relations are
\begin{equation}\label{comm-6-dim-c20}
\begin{split}
&[v_1, v_2]=v_1-3c_1 v_4+c_0v_6, \quad [v_1, v_3]=2v_2-\frac{v_6}{2}, \quad [v_1, v_4]=v_5,\\
&[v_1, v_5]=c_1 v_6, \quad [v_2, v_3]=v_3, \quad [v_2, v_4]=\frac{1}{2}v_4, \quad [v_2, v_5]=-\frac{1}{2}v_5,\\
&[v_3, v_5]=-v_4,  \quad [v_4, v_5]=\frac{v_6}{2}.
\end{split}
\end{equation}

\item $c_2=-\kappa^2$, $\kappa>0$.
\begin{equation}\label{6-dim-c2n}
\begin{split}
&v_1=\gen t,\\
&v_2=e^{4\kappa t}\gen t+\frac{1}{\kappa} e^{4\kappa t}\sqrt{a}(-c_1+2\kappa^2 I)\gen x-\\
&\quad \frac{1}{4\kappa^2}e^{4\kappa t}[c_1^2-4c_0\kappa^2+4\kappa^3+8\kappa^4I^2+2c_1 \kappa J-4\kappa^2I(2c_1+\kappa J)]u\gen u,\\
&v_3=e^{-4\kappa t}\gen t+\frac{1}{\kappa} e^{-4\kappa t}\sqrt{a}(c_1-2\kappa^2 I)\gen x-\\
&\quad \frac{1}{4\kappa^2}e^{-4\kappa t}[c_1^2-4c_0\kappa^2-4\kappa^3+8\kappa^4I^2-2c_1 \kappa J+4\kappa^2I(-2c_1+\kappa J)]u\gen u,\\
&v_4=e^{2\kappa t}\sqrt{a}\gen x+\frac{1}{2\kappa}e^{2\kappa t}[c_1-2\kappa^2 I+\kappa J]u\gen u,\\
&v_5=e^{-2\kappa t}\sqrt{a}\gen x+\frac{1}{2\kappa}e^{-2\kappa t}[-c_1+2\kappa^2 I+\kappa J]u\gen u,\\
&v_6=u\gen u.
\end{split}
\end{equation}
The non-zero commutation relations are
\begin{equation}\label{comm-6-dim-c2n}
\begin{split}
&[v_1, v_2]=4\kappa v_2, \quad [v_1, v_3]=-4\kappa v_3, \quad [v_1, v_4]=2\kappa v_4,\\
&[v_1, v_5]=-2\kappa v_5, \quad [v_2, v_3]=-2\kappa (4v_1+r v_6), \quad [v_2, v_5]=-4\kappa v_4, \\
&[v_3, v_4]=4\kappa v_5,  \quad [v_4, v_5]=2 \kappa v_6,
\end{split}
\end{equation}
where $r=4c_0+(c_1/\kappa)^2$.

\item $c_2=\kappa^2$, $\kappa>0$.
\begin{equation}\label{6-dim-c2p}
\begin{split}
&v_1=\gen t,\\
&v_2=\cos{4\kappa t}\gen t-\frac{1}{\kappa}\sin{4\kappa t} \sqrt{a}(c_1+2\kappa^2 I)\gen x+\\
&\quad [\cos{4\kappa t}(c_0+\frac{c_1^2}{4\kappa^2}+2(c_1+\kappa^2I)I)+\sin{4\kappa t}(\kappa-\frac{1}{2\kappa}(c_1+2\kappa^2I)J)]u\gen u,\\
&v_3=\sin{4\kappa t}\gen t+\frac{1}{\kappa}\cos{4\kappa t} \sqrt{a}(c_1+2\kappa^2 I)\gen x+\\
&\quad [\cos{4\kappa t}(-\kappa+\frac{1}{2\kappa}(c_1+2\kappa^2I)J)+\sin{4\kappa t}(c_0+\frac{c_1^2}{4\kappa^2}+2(c_1+\kappa^2I)I)]u\gen u,\\
&v_4=\cos{2\kappa t}\sqrt{a}\gen x+\frac{1}{2\kappa}[\sin{2\kappa t}(c_1+2\kappa^2 I)+\kappa \cos{2\kappa t}J]u\gen u,\\
&v_5=\sin{2\kappa t}\sqrt{a}\gen x+\frac{1}{2\kappa}[-\cos{2\kappa t}(c_1+2\kappa^2 I)+\kappa\sin{2\kappa t} J]u\gen u,\\
&v_6=u\gen u.
\end{split}
\end{equation}
The non-zero commutation relations are
\begin{equation}\label{comm-6-dim-c2p}
\begin{split}
&[v_1, v_2]=-4\kappa v_3, \quad [v_1, v_3]=4\kappa v_2, \quad [v_1, v_4]=-2\kappa v_5,\\
&[v_1, v_5]=2\kappa v_4, \quad [v_2, v_3]=\kappa (4v_1+s v_6), \quad [v_2, v_4]=2\kappa v_5, \quad [v_2, v_5]=2\kappa v_4, \\
&[v_3, v_4]=-2\kappa v_4, \quad [v_3, v_5]=2\kappa v_5,  \quad [v_4, v_5]=- \kappa v_6,
\end{split}
\end{equation}
where $s=4c_0 -(c_1/\kappa)^2$.

\end{enumerate}
The symmetry algebras in cases \eqref{6-dim-c20}, \eqref{6-dim-c2n} and  \eqref{6-dim-c2p} has  a Levi-decom\-po\-si\-tion structure in the form of semi-direct sum $\lie_6=\Sl(2,\mathbb{R})\oplus_s \nil(1)$, where $\nil(1)$ is the three-dimensional nilradical (Heisenberg algebra) with $v_6$ being the center element. $\lie_6$ is isomorphic to the heat algebra. A compete study of the symmetry properties of the heat equation has recently been discussed in \cite{KovalPopovych2023}. A general treatment of Kolmogorov and Fokker--Planck type PDEs from a new  perspective   can be found in \cite{Opanasenko2022}.

We note that the vector field $v_3$ has a sign error in the two exponential terms of  \cite{Guengoer2018} and the basis elements \eqref{6-dim-c2p} appears mistakenly as a copy of \eqref{6-dim-c2n} in the same reference.

Because of linearity, the solution superposition  symmetries $w(x,t)\gen u$, where $w$ solves \eqref{mainLPE}, form an infinite-dimensional ideal. After this ideal is excluded, a generic spatially dependent equation may have only the two-dimensional kernel algebra generated by
\begin{equation}\label{ess-sym}
  \gen t,   \quad u\gen u.
\end{equation}

We summarise the above results of Sections \ref{Subsection4-dim} and  \ref{Subsection6-dim} as a theorem. It characterizes the subclasses possessing four- and six-dimensional
essential algebras.

\begin{thm}\label{sym-algebra}
The dimension of the non-trivial  symmetry algebra  of Eq. \eqref{mainLPE}  is at least four. A four-dimensional symmetry algebra occurs if and only if
\begin{equation}\label{L4}
  K(x)=f_4(I):= \frac{\mu}{I^2}+c_2 I^2+c_0,\quad \mu\ne 0,
\end{equation}
a six-dimensional one (maximal) if and only if
\begin{equation}\label{L6}
  K(x)=f_6(I):= c_2 I^2+c_1 I+c_0
\end{equation}
for some constants $c_2$, $c_1$, $c_0$, $\mu$.
The first algebra   is locally isomorphic to $\lie_4=\Sl(2,\mathbb{R})\oplus \mathbb{R}$, the second one  isomorphic to the Schr\"odinger (or heat) algebra.
\end{thm}

\subsection{Discussion of solutions of the symmetry criteria}
Regarding $I$ as the independent variable in the definition of $K$ we can write the invariance condition $K(x)=f(I)$, where $f\in\{f_4,f_6\}$ in the form
\begin{equation}\label{Riccati-0}
\frac{1}{2}\frac{d\Omega}{dI}-\frac{1}{4}\Omega^2+\frac{q}{a}(a'-b+q)=f(I),  \quad q'=c, \quad \Omega=J+\frac{2q}{\sqrt{a}}.
\end{equation}
This ODE can be integrated as a Riccati equation if the last term on the left hand side of \eqref{Riccati-0} is expressed in terms of a function of $I$ alone.
In the simpler case where Eq. \eqref{mainLPE} is in conserved form ($a''-b'+c=0$),  it becomes tractable in terms of special functions. The conserved PDE has the form
\begin{equation}\label{conservedform}
  u_t=\partial_x[au_x+(b-a')u].
\end{equation}
Letting $a=p$, $q=b-a'$ leads to the Fokker--Planck equation
\begin{equation}\label{FP}
  u_t=\partial_x(pu_x+qu)
\end{equation}
with  $J=-[(\sqrt{p})'+q/\sqrt{p}]$.
In this case, in \eqref{Riccati-0} we have $a'-b+q=\text{const.}$ which we pick zero for simplicity.  Thus $\Omega(I)$ satisfies the special Riccati equation
\begin{equation}\label{Riccati-1}
  \frac{1}{2}\frac{d\Omega}{dI}-\frac{1}{4}\Omega^2=f(I),
\end{equation}
where
\begin{equation}\label{Q}
  \Omega=-(\sqrt{p})'+\frac{q}{\sqrt{p}}.
\end{equation}

From \eqref{Q} we find that for given coefficient $p$ solving \eqref{Riccati-1} we can determine $q$ such that the main PDE admits a non-trivial symmetry group
\begin{equation}\label{p-q}
  q=\frac{p'}{2}+\sqrt{p}\Omega(I).
\end{equation}
The solutions of \eqref{Riccati-1} can be expressed in terms of the confluent hypergeometric functions. The Cole--Hopf transformation $\Omega=-2v'/v$ linearizes \eqref{Riccati-1} to the second order  ordinary differential equation (ODE)
\begin{equation}\label{confl-0}
  4v''-fv=0.
\end{equation}
Here  we have replaced $f$ by $-(1/4)f$ for convenience.

When $f=f_6$ the transformation of both the independent and dependent variables
\begin{equation}\label{tr1}
  v=e^{-z/2}w(z),  \quad z=\frac{\sqrt{c_2}}{2}\left(I+\frac{c_1}{2c_2}\right)^2
\end{equation}
maps this ODE to confluent hypergeometric (Kummer) equation
\begin{equation}\label{hyper-conf-6}
 zw''+\left(\frac{1}{2}-z\right)-aw=0,
\end{equation}
where
$$a=\frac{1}{4}\left[1+\frac{4c_0c_2-c_1^2}{8c_2^{3/2}}\right].$$

For the case $f=f_4$  we apply the transformation
\begin{equation}\label{tr2}
  v=z^{s}e^{-z/2}w(z),  \quad z=\frac{\sqrt{c_2}}{2}I^2,   \quad s=\frac{1}{2}(1\pm \sqrt{1+\mu})
\end{equation}
to transform \eqref{confl-0} to another confluent hypergeometric (Kummer) differential equation
\begin{equation}\label{conf-hyper-4}
  zw''+(2s-z)w'-\left(\frac{c_0}{8\sqrt{c_2}}+\frac{1}{4}+\frac{s}{2}\right)w=0.
\end{equation}

\section{Applications}\label{app}
In this section we illustrate the ideas developed in Section \ref{point-symm} with  several  PDEs with varying diffusion and drift coefficients arising in a wide range of applications.  Our intention is to show how  they can serve to effectively simplify the symmetry group analysis, which otherwise might be rather tricky,  specifically when at least one coefficient, say the drift term, is allowed to remain  arbitrary (see Example \ref{volat}).

\begin{example}\label{t-depen-drift}
We first consider a PDE  whose  drift term  has  both space and time dependence \begin{equation}\label{pde-xt}
  a=1, \quad b=m(t)+n(t)x, \quad c=q(t)+r(t)x.
\end{equation}
We first let $n\ne 0$. We impose the criterion \eqref{non-trivial-sym-cond}  for the PDE to be transformable to the heat equation
$$K(x,t)=c_2(t)I^2+c_1(t)I+c_0(t)$$ with $K$ as given by \eqref{K-reduced},
more explicitly
$$\frac{1}{2}J_x-\frac{1}{4}J^2+c+\frac{1}{2}\int (\dot{m}+\dot{n}x)dx =c_2(t)x^2+c_1(t)x+c_0(t),   \quad J(x,t)=-b(x,t)$$
for some functions $c_2, c_1, c_0$ that need to be fixed conveniently. This occurs if and only if
\begin{equation}\label{conds-m-n}
  \dot{n}-n^2=4c_2(t),  \quad \dot{m}-mn+2r=2c_1(t),  \quad q-\frac{n}{2}-\frac{m^2}{4}=c_0(t).
\end{equation}
With this choice the symmetry algebra of the corresponding PDE should be isomorphic to that of the heat equation.

Infinitesimal generators $\xi, \phi$ of the general symmetry vector field \eqref{VF} are    given by
\begin{equation}\label{xxi}
  \xi(x,t)=\frac{1}{2}\dot{\tau}x+\rho(t),
\end{equation}
and
\begin{equation}\label{phi-2}
  \phi(x,t)=-\frac{1}{8}\ddot{\tau}x^2-\frac{1}{2}\dot{\rho}x-\left(\frac{1}{4}\dot{\tau} x+\frac{1}{2}\rho\right)b(x,t) -\frac{\tau}{2}\int {b_t}dx+\sigma(t),
\end{equation}
where $\tau, \rho, \sigma$ satisfy a classifying equation splitting into the set of ODEs
(compare with \eqref{deteqs})
\begin{equation}\label{tau-rho-sigma}
\begin{split}
   & \dddot{\tau}+16c_2\dot{\tau}+8\dot{c_2}\tau=0,  \quad \ddot{\rho}+4c_2\rho+3c_1\dot{\tau}+2\dot{c_1}\tau=0 \\
    & \dot{\sigma}=\frac{d}{dt}(c_0\tau)+c_1\rho-\frac{\ddot{\tau}}{4}.
\end{split}
\end{equation}
The function $\tau(t)$ has the fundamental system of solutions $\curl{\psi_1^2,\psi_1\psi_2,\psi_2^2}$, where $\psi_1, \psi_2$ are two independent solutions of the ODE $\psi''+4c_2\psi=0$.

For fixed values of $c_2(t)$ in the first condition of \eqref{conds-m-n}, we can specify $n(t)$ by solving a Riccati equation. Otherwise, for given $n(t)$,  $c_2(t)$ remains fixed.
Hence, fixing $c_2(t)=\varepsilon\lambda^2/4$, $\lambda=\text{const.}$ leads to the Riccati equation $\dot{n}-n^2=\varepsilon \lambda^2$, $\varepsilon\in\curl{0,1,-1}$ with solutions in terms of elementary functions.  $T$ is found as a solution of the Schwarzian equation $\curl{T,t}=2\varepsilon \lambda^2$. Once $n$ has been determined, then  any solution $\omega$ of the  ODE
\begin{equation}\label{omega}
  \ddot{\omega}+\varepsilon \lambda^2 \omega=2c_1=\dot{m}-mn+2r
\end{equation}
will complete the required transformation  \eqref{ToHeatTr}.

Moreover, setting $c_2(t)=(A/4)t^{\alpha}$ for $\alpha=-2$ and $\alpha=-4k/(2k\pm 1)$, $k\in \mathbb{N}$  gives an explicitly solvable Riccati equation  of which solutions will produce many time-dependent   drift coefficients $n(t)$ in terms of elementary functions. Solutions of $\curl{T,t}=2A t^{\alpha}$ will  also be elementary. Both differential equations will require solving the second order linear ODE $\ddot{z}+A t^{\alpha}z=0$.

For example, for $A=1$, $\alpha=-4$, $c_2(t)=t^{-4}/4$ we find
$$n(t)=t^{-2}(\cot(t^{-1}+C_0)-t),  \quad T(t)=\frac{\alpha \cos(\frac{1}{t})+\beta \sin(\frac{1}{t})}{\gamma \cos(\frac{1}{t})+\delta \sin(\frac{1}{t})}, \quad \alpha\delta-\beta\gamma=1.$$
For the particular choice $m=0$, $q=0$ and $r=r_0t^{-3}$, we have $c_1(t)=r(t)$, $c_0(t)=-n(t)/2$ we can solve the system \eqref{tau-rho-sigma}
$$\tau(t)=t^2\left[ \tau_1\cos^2 \left(\frac{1}{t}\right)+\tau_2\sin \left(\frac{2}{t}\right)+\tau_3\sin^2 \left(\frac{1}{t}\right)\right],$$
$$\rho(t)=t\left[ \rho_1 \cos \left(\frac{1}{t}\right)+\rho_2 \sin \left(\frac{1}{t}\right)\right]+r_0t\left[(\tau_1-\tau_3)\sin \left(\frac{2}{t}\right)-2\tau_2 \cos \left(\frac{2}{t}\right)\right].$$ Finally, we obtain $$\sigma=c_0\tau+r_0\int t^{-3}\rho(t)-\frac{\dot{\tau}}{4}+\sigma_0,$$
whose explicit form will not be produced here. Six arbitrary integration constants enter into the infinitesimal symmetries.

In summary, we have constructed a PDE of the form
\begin{equation}\label{pde}
  u_t=u_{xx}+\frac{x}{t^2}\left[\cot\left(\frac{1}{t}+C_0\right)-t\right]u_x+r_0t^{-3}xu
\end{equation}
admitting a symmetry algebra isomorphic to that of the heat equation with a basis given as discussed above. The related point transformation is again easy to write down from the formula \eqref{ToHeatTr} in which $\omega(t)$ can be taken as a private solution $\omega=2r_0t$  of $\ddot{\omega}+t^{-4}\omega=2r_0t^{-3}$.

The  case $n=0$ ($c_2=0$) is  easy to handle and we find the corresponding simplest form of the transformation  as
\begin{equation}\label{trans-space-indep}
  \tilde{t}=t,  \quad \tilde{x}=x+\int (m+2s)dt,  \quad u=C\exp\left[sx+\int [s(s+m)+ q]dt\right]\tilde{u},
\end{equation}
where $r=\dot{s}$.

Another simple case occurs when  $m=0$,  $n=-1/t$ ($\varepsilon=0$) and $r=0$ then we have $c_2=c_1=0$ and $c_0(t)=q(t)+1/(2t)$ and the corresponding transformation depending on four arbitrary parameters becomes
\begin{equation}\label{tr-m-zero}
\begin{split}
   & \tilde{t}=M(t)=\frac{\alpha t+\beta}{\gamma t+\delta},  \quad \tilde{x}=\frac{\sqrt{\Delta}}{\gamma t+\delta}(x+t), \quad \Delta=\alpha \delta-\beta \gamma, \\
    & u=C(\gamma t+\delta)^{-1/2}\sqrt{t}\exp\left[\frac{(x+t)^2}{4t}-\frac{\gamma(x+t)^2}{4(\gamma t+\delta)}+\int q(t)dt\right]\tilde{u}(\tilde{x},\tilde{t}),
\end{split}
\end{equation}
where we have chosen $\omega(t)=t$ in \eqref{ToHeatTr}. From this group action on solutions, using the trivial solution $\tilde{u}=1$ we obtain the non-trivial solution of the diffusion equation
\begin{equation}\label{spec-diff}
  u_t=u_{xx}-\frac{x}{t}u_x+q(t)u=0
\end{equation}
as
\begin{equation}\label{sol-diff-xt}
  u=C(\gamma t+\delta)^{-1/2}\sqrt{t}\exp\left[\frac{\delta(x+t)^2}{4t(\gamma t+\delta)}+\int q(t)dt\right].
\end{equation}
If $r\ne 0$,  $\omega$ should be taken as any solution of $\ddot{\omega}=2r$.

By solving the system \eqref{tau-rho-sigma} it can be shown that he finite dimensional part of the Lie symmetry algebra of \eqref{spec-diff} for $q=0$ is spanned by
\begin{equation}\label{6-dim-sym}
\begin{split}
&v_1=\gen t+\left(\frac{1}{2t}-\frac{x^2}{4t^2}\right)u\gen u,\\
&v_2=t\gen t+\frac{x}{2}\gen x,\\
&v_3=t^2\gen t+xt \gen x,\\
&v_4=t\gen x,\\
&v_5=t^2\gen t+(1+xt)\gen x+\frac{x}{2t}u\gen u,\\
&v_6=u\gen u.
\end{split}
\end{equation}

The case $\varepsilon=-1$: We have  $n(t)=-\lambda \tanh \lambda t$. We can single out from \eqref{conds-m-n} another time-space dependent invariant PDE of the form \eqref{pde-xt} with $m, r$ fixed further
$$m(t)=t\sech \lambda t,  \quad r(t)=-\frac{1}{2}\sech \lambda t.$$ Subsequently, functions $c_2(t), c_1(t), c_0(t)$ are also fixed as
$$c_2=-\frac{\lambda^2}{4}, \quad c_1=0, \quad c_0=q(t)+\frac{\lambda}{2}\tanh \lambda t-\frac{1}{4}t^2 \sech^2 \lambda t.$$
For $q=0$, a basis for the symmetry algebra is immediately obtained from \eqref{xxi}-\eqref{phi-2}
by solving the system \eqref{tau-rho-sigma} in the form
\begin{equation}\label{tau-rho-sigma-solved}
  \begin{split}
   & \tau=\tau_0+\tau_1 \cosh 2\lambda t+\tau_2 \sinh 2\lambda t, \quad \rho=\rho_1 \cosh\lambda t+\rho_2 \sinh\lambda t, \\
   & \sigma=-\frac{1}{4}\sech^2 \lambda t[(\tau_0+\tau_1 \cosh 2\lambda t+\tau_2 \sinh 2\lambda t) t^2+(\tau_1-\tau_0)\lambda\sinh 2\lambda t]+\sigma_0.
\end{split}
\end{equation}
The arbitrary constants $\tau_0, \tau_1, \tau_2$ correspond to  $\Sl(2,\mathbb{R})$ part and $\rho_1,\rho_2,\sigma_0$ to   $\nil(1)$ Heisenberg part  of the maximal symmetry algebra.   We already know that this algebra is isomorphic to that of the heat equation. Of course, the transformation mapping \eqref{pde-xt} to the heat equation is given by \eqref{ToHeatTr}.

The other  possibility $\varepsilon= 1$ can be dealt with in the same manner.

\end{example}

\begin{rmk}
  We comment that the complete characterization of the drift functions for the existence of a non-trivial Lie point algebra of symmetries   of a class of  Fokker--Planck equations associated to Ito integrable equations   analyzed in the recent paper \cite{GaetaRodriguez2023} is readily recovered from the results of our above discussion of Example \ref{t-depen-drift} in the special case $q(t)=n(t)$, $r(t)=0$.

The following transformation mapping this  PDE in conserved form  to    the heat equation given in \cite{JohnpillaiMahomed2001} is surely covered in our general transformation formula \eqref{ToHeatTr}
\begin{equation}\label{FP-heat}
  \tilde{t}=T(t),  \quad \tilde{x}=e^{\alpha(t)}x+\beta(t),  \quad u=e^{\alpha(t)}\tilde{u},
\end{equation}
where $\alpha,\beta$ satisfy
\begin{equation}\label{T-al-be}
  \dot{T}=e^{2\alpha(t)},\quad \dot{\alpha}=n,  \quad \dot{\beta}=e^{\alpha(t)}m.
\end{equation}
In order to achieve this we choose $\beta=e^{\alpha}\omega$ and observe that
\begin{equation}\label{T-omega}
\curl{T,t}=2(\ddot{\alpha}-\dot{\alpha}^2)=2(\dot{n}-n^2),  \quad \dot{\omega}+n \omega=m,  \quad \ddot{\omega}+(\dot{n}-n^2)\omega=\dot{m}-mn.
\end{equation}
This means then the pair of ODEs  \eqref{ODE-pair} is exactly the set of constraints  \eqref{conds-m-n} with $r=0$ and if we further choose $c_0=(2n-m^2)/4$, the whole term multiplying $\tilde{u}$ in the formula \eqref{ToHeatTr} is independent of $x$ and the rest boils down  to $e^{\alpha}$. So we arrive at \eqref{FP-heat}-\eqref{T-al-be}.

We note that in the general case \eqref{pde-xt},  a  transformation similar to \eqref{FP-heat} exists in the form
\begin{equation}\label{non-FP-heat}
\tilde{t}=\int e^{2\alpha(t)}dt,  \quad \tilde{x}=e^{\alpha(t)}x+\beta(t),  \quad u=\exp\left[e^{\alpha(t)}\gamma(t) x+A(t)\right]\tilde{u},
\end{equation}
where
\begin{equation}\label{gama}
  \dot{\alpha}=n,  \quad \dot{\beta}=e^{\alpha}(m+2e^{\alpha}\gamma), \quad \gamma(t)=\int e^{-\alpha(t)}r(t)dt
\end{equation}
and
$$A(t)=\int^t\Big(q(s)+e^{2\alpha(s)}\gamma^2(s)
+m(s)e^{\alpha(s)}\gamma(s)\Big)ds.$$
Moreover, $\beta$ satisfies $\ddot{\beta}-2n\dot{\beta}=2e^{\alpha}c_1$.
The Fokker--Planck case means we have $q=n=\dot{\alpha}$ and $\gamma=0$ and the transformation \eqref{non-FP-heat}-\eqref{gama} simplifies to \eqref{FP-heat}-\eqref{T-al-be}.

For the special PDE \eqref{pde} ($C_0=0$) with the help of the following functions
$$e^{2\alpha}=t^{-2}\csc^2(t^{-1}),  \quad \beta=2r_0\csc(t^{-1}),  \quad \gamma=r_0\cos(t^{-1})$$
the  transformation \eqref{non-FP-heat} specializes to
\begin{equation}\label{tr-pde}
\begin{split}
     \tilde{t}&=\cot(t^{-1}),\quad \tilde{x}=\csc(t^{-1})(t^{-1}x+2r_0), \\
      u&=\exp\left[\frac{r_0x}{t}\cot\frac1t
+ r_0^2\left(t^{-1}+\cot(t^{-1})\right)\right]\tilde{u},
\end{split}
\end{equation}
which provides us with a mapping of \eqref{pde} to the heat equation.

A final note is that if we additionally let the coefficient $c(x,t)$ to be a quadratic polynomial in $x$ as $c=q(t)+r(t)x+s(t)x^2$, then the corresponding PDE still remains  transformable to the heat equation. However, there is no such simple formula like \eqref{non-FP-heat}-\eqref{gama}. In this case, only the first relation of \eqref{conds-m-n} changes as $\dot{n}-n^2+4s=4c_2$.

\end{rmk}

\begin{example}\label{volat}
The following PDE as local volatility model in mathematical finance has been studied from the symmetry point of view (see \cite{CraddockGrasselli2020} for the details):
\begin{equation}\label{CraddockGrasselli}
  u_t=\frac{1}{2}\sigma^2(x) u_{xx}-g(x)u,  \quad \sigma>0,
\end{equation}
where $\sigma$ and $g$ are functions defined in some subinterval of $\mathbb{R}$ and the drift coefficient is zero.

All the symmetry calculation results of this paper can immediately be obtained  from our general result for the given coefficients of  PDE \eqref{CraddockGrasselli}. All that is needed is the calculation of the semi-invariant $K(x)$ defined by \eqref{semi-inv}. Then, the condition for the equation to have a six-dimensional symmetry algebra is simply obtained by forming the condition (see formula \eqref{K-2})
\begin{equation}\label{criterium}
  K(x)=c_2 I^2(x)+c_1 I(x)+c_0,
\end{equation}
where $c_i$, $i=0,1,2$ are some constants to be determined later. The coefficients of the symmetry vector field are given by \eqref{xi} and \eqref{phi}
under the condition \eqref{comp}. Here there is a slight change of the factor $\rho(t)$: $\rho\to\rho/\sqrt{2}$. All the infinitesimal generators of the symmetry algebra can be produced using the formulae  of Subsection \eqref{Subsection6-dim}.

In this case,
$$a(x)=\frac{1}{2}\sigma(x)^2, \quad b(x)=0, \quad c(x)=-g(x)$$ and it is straightforward to calculate
\begin{equation}\label{auxil}
  I(x)=\sqrt{2} \psi(x),  \quad J(x)=\frac{\sigma'(x)}{\sqrt{2}},  \quad K(x)=\frac{1}{4}\sigma \sigma''-\frac{1}{8}\sigma'^2-g,
\end{equation}
where $\psi(x)$ denotes the Lamperti transformation implicitly defined by $\psi'=1/\sigma(x)$.
Condition \eqref{criterium} now gives rise to
\begin{equation}\label{2ndODE}
  \frac{1}{8}(2\sigma \sigma''-\sigma'^2)-g=2c_2 \psi^2+c_1\sqrt{2}\psi+c_0.
\end{equation}

This condition, when differentiated once,
\begin{equation}\label{Kprime}
  K'(x)=\frac{1}{4}\sigma \sigma'''-g'=\frac{4c_2 \psi}{\sigma}+\frac{\sqrt{2}c_1}{\sigma}
\end{equation}
coincides with the result of  Theorem (3.4) of \cite{CraddockGrasselli2020} for $c_1=A/\sqrt{2}$, $c_2=B/4$.
Also,  Theorems 3.2, 3.4, 3.5 of the same work contain calculations that can  be easily obtained in terms of $I, J$ without having to start from scratch with applying  Lie's method for different coefficients of the equation. For the Lie algebra given in Theorem 3.5, it is sufficient to choose $c_0=0$ in \eqref{6-dim-c20}. We note a misprint in $v_2$: the factor $1/8$ of the last term of the $u$-coefficient should be $1/4$.

An interesting observation here is that condition \eqref{2ndODE}, when the set of functions $(c_2,c_1,c_0)$ are eliminated by three successive differentiations with respect to $x$, turns into a  nonlinear  ODE of fifth order in $\sigma(x)$  and fully agrees with the singular invariant equation $\lambda=0$ given in\cite{JohnpillaiMahomed2001}.

By \eqref{2ndODE}, if we allow $\sigma$ to be a fixed quadratic normal volatility model
\begin{equation}\label{quadra-volat}
  \sigma(x)=\alpha+\beta x+\gamma x^2, \quad \gamma\ne 0,
\end{equation}
then  $g(x)$ can be determined as a quadratic polynomial in $\psi(x)$
$$g=g_2 \psi^2+g_1\psi+g_0,$$
where $g_2=-2c_2$, $g_1=-\sqrt{2}c_1$, $g_0=-(c_0+1/8(\beta^2-4\alpha\gamma))$.

For the  choice  $c_1=c_2=0$, $c_0=-1/8(\beta^2-4\alpha\gamma)$ we have $g=0$. This implies that if the quadratic \eqref{quadra-volat} has distinct roots ($c_0\ne 0$), then the vector fields $v_2$ and $v_3$ of \eqref{6-dim-c20} ($v_2$ and $v_1$ with the notation of \cite{CraddockGrasselli2020})  will also depend on $c_0$. So in that case $v_1$ and $v_2$ of \cite{CraddockGrasselli2020}  are no longer correct and should be rectified.

If $\sigma$ has two equal roots then $c_0=0$.
When the quadratic $\sigma$ \eqref{quadra-volat} has two distinct and multiple real roots two different fundamental solutions  have been obtained in \cite{CraddockGrasselli2020} by using the  action of the  symmetry $v_5$ (free of $c_0$) ($v_4$ with the notation of \cite{CraddockGrasselli2020}) on the constant solution $u=1$.

If we choose $\sigma=\sqrt{2}(1+k^2x^2)$ ($c_0=k^2$) then this process gives a PDE that arises in a study of Brownian motion
\begin{equation}\label{Brownian}
  u_t=(1+k^2 x^2)^2u_{xx}.
\end{equation}

The coefficients are  $a(x)=(1+k^2x^2)^2$, $b(x)=c(x)=0$ and we find
$$I(x)=\frac{1}{k}\arctan (kx), \quad J(x)=2k^2x,  \quad K(x)=k^2.$$
We check that condition \eqref{non-trivial-sym-cond} for the PDE \eqref{Brownian} to have maximal  symmetry is satisfied for $c_1=c_2=0$ and $c_0=k^2$. It follows that a reduction to the first canonical form (heat equation) is possible by the point transformation \eqref{ToHeatTr}  (for the simple choice $\omega=0$) given by
\begin{equation}\label{heat-tr-1}
\begin{split}
&\tilde{t}=M(t)=\frac{\alpha t+\beta}{\gamma t+\delta},  \quad \tilde{x}=\frac{1}{k}\sqrt{\Delta}(\gamma t+\delta)^{-1}\arctan kx, \\
&u=C (\gamma t+\delta)^{-1/2}(1+k^2x^2)^{1/2}\exp\left[-\frac{\gamma(\arctan kx)^2}{4k^2(\gamma t+\delta)}+k^2 t\right]\tilde{u},
\end{split}
\end{equation}
where $C$ is a nonzero constant.
As a bonus, the transformation formula \eqref{heat-tr-1} in the limit $k\to 0$ produces the well-known Appell's transformation mapping the heat equation into each other \cite{Appell1982}
\begin{equation}\label{appell}
\begin{split}
&\tilde{t}=M(t)=\frac{\alpha t+\beta}{\gamma t+\delta},  \quad \tilde{x}=\sqrt{\Delta}(\gamma t+\delta)^{-1}x, \\
&u=C |\gamma t+\delta|^{-1/2}\exp\left[-\frac{\gamma x^2}{4(\gamma t+\delta)}\right]\tilde{u}(\tilde{x},M(t)).
\end{split}
\end{equation}
A well-known fact is that transformation \eqref{appell} maps the constant solution $\tilde{u}=1$ to the heat kernel (fundamental solution of the heat equation) up to a multiplicative constant. This argument has substantially been used in the present work to produce physically significant solutions.

From \eqref{6-dim-c20} we find that the corresponding six-dimensional symmetry algebra is spanned by the vector fields
\begin{equation}\label{6-dim-basis}
\begin{split}
&v_1=\gen t,\\
&v_2=t\gen t+\frac{1}{2k}(1+k^2x^2)\arctan kx\gen x+\frac{k}{2}(2kt+x \arctan kx)u\gen u,\\
&v_3=t^2\gen t+\frac{t\arctan kx}{k}(1+k^2x^2)\gen x-\frac{1}{4}[2t-4k^2t^2-4ktx \arctan kx+\\
& \qquad +\frac{(\arctan kx)^2}{k^2}]u\gen u,\\
&v_4=t(1+k^2x^2)\gen x+[k^2tx-\frac{\arctan kx}{2k}]u\gen u,\\
&v_5=(1+k^2x^2)\gen x+k^2 x u\gen u,\\
&v_6=u\gen u.
\end{split}
\end{equation}
From the non-zero commutation relations  of \eqref{comm-6-dim-c20} we immediately have
\begin{equation}\label{comm}
\begin{split}
&[v_1, v_2]=v_1+k^2v_6, \quad [v_1, v_3]=2v_2-\frac{v_6}{2}, \quad [v_1, v_4]=v_5,\\
& [v_2, v_3]=v_3, \quad [v_2, v_4]=\frac{1}{2}v_4, \quad [v_2, v_5]=-\frac{1}{2}v_5,\\
&[v_3, v_5]=-v_4,  \quad [v_4, v_5]=\frac{v_6}{2}.
\end{split}
\end{equation}
In the limit $k\to 0$ ($a\to 1$ and $c_0\to 0$), the symmetry algebra spanned by \eqref{6-dim-c20}  reduces to that of the standard heat equation by a slight change of the  basis $v_2\to 2v_2-v_6/2$.


Finally, we remark that with $c_1=c_2=0$ and $g=-h(x)\sigma^2/2$ for some function $h$, our PDE  \eqref{CraddockGrasselli} becomes
\begin{equation}\label{quadra-volat-2}
  u_t=\frac{\sigma^2}{2}(u_{xx}+hu)
\end{equation}
and \eqref{2ndODE} has the form
\begin{equation}\label{2ndODE-2}
  \frac{1}{4}(2\sigma \sigma''-\sigma'^2)+h\sigma^2=2c_0.
\end{equation}
General solution of \eqref{2ndODE-2} is given by
\begin{equation}\label{sol-sigma}
  \sigma=\alpha \varphi_1^2+\beta \varphi_1\varphi_2+\gamma \varphi_2^2, \quad (4\alpha\gamma-\beta^2)W^2=8c_0
\end{equation}
where the set $\curl{\varphi_1, \varphi_2}$ is the fundamental set of solutions of the linear ODE $\varphi''+h\varphi=0$ and $W$ is its Wronskian \cite{GuengoerTorres2017}. Notice that $\varphi$ coincides with the stationary solution of \eqref{quadra-volat-2}.

We recall that Eq. \eqref{2ndODE-2} for given $p(x)$ and $c_0$ is connected with the Ermakov equation
\begin{equation}\label{ermakov}
  \chi''+h \chi=2c_0 \chi^{-3}
\end{equation}
through the transformation $\sigma=\chi^2$. We refer to \cite{GuengoerTorres2017} for more details of this equation and its extensions.

In particular, if we set $h=0$ ($g=0$), then the fundamental set is $\curl{1,x}$ with $W=1$ and we recover the quadratic volatility \eqref{quadra-volat} under the condition $(\alpha\gamma-4\beta^2)=2c_0$ as obtained above.

Using one-dimensional subalgebras of \eqref{6-dim-c20} interesting group invariant solutions of \eqref{quadra-volat-2} with $h=0$ can be obtained. We recall that the vector fields $(v_1,v_2,v_3)$ span an algebra $\ort(2,1)$ of pseudo-rotations, up to change of basis, isomorphic to the $\Sl(2,\mathbb{R})$ algebra.
For example,  solution invariant under the subalgebra $v_4$ (up to a multiplicative constant) of \eqref{6-dim-c20}
$$v_4=2t\sigma\gen x+(t\sigma'-2\psi)u\gen u$$
should have the form
\begin{equation}\label{inv-sol-0}
  u=R(t)\sigma^{1/2}\exp\left[-\frac{\psi^2}{2t}\right].
\end{equation}
Substituting \eqref{inv-sol-0} into \eqref{quadra-volat-2} and using the relation \eqref{2ndODE-2} we find
$$\frac{\dot{R}}{R}=c_0-\frac{1}{2t}$$ leading to the solution
\begin{equation}\label{inv-sol-v4}
  u(x,t)=C\sqrt{\frac{\sigma}{t}}\exp\left[c_0t-\frac{\psi^2}{2t}\right].
\end{equation}
This solution includes the heat kernel for $\sigma=\sqrt{2}$ and $\psi=x/\sqrt{2}$ ($c_0=0$). A case where $c_0\ne 0$ arises, among many others, is $\sigma=\sin \nu x \cos \nu x$ ($h=\nu^2\ne 0$, $c_0=-\nu^2/8$) and $\psi=\nu^{-1}\ln(\tan \nu x)$. Then solution \eqref{inv-sol-v4} specializes to
$$u=Ct^{-1/2}(\sin 2\nu x)^{1/2}\exp\left[-\frac{\nu^2 t}{8}-\frac{(\ln(\tan \nu x))^2}{2\nu^2 t}\right].$$

In a similar way, invariance under the symmetry
$$v_5=\sigma\gen x+\frac{1}{2}\sigma' u\gen u$$
produces the solution
\begin{equation}\label{inv-sol-2}
  u(x,t)=C\sqrt{\sigma}e^{c_0t}.
\end{equation}
Both invariant solutions \eqref{inv-sol-v4} and \eqref{inv-sol-2} could be obtained from \eqref{ToHeatTr} by transforming the constant solution $\tilde{u}=1$ into $u(x,t)$ by choosing $T(t)=-1/t$ and $T(t)=t$, respectively. Moreover, in the special case $\sigma=\gamma(x-r)^2$ ($c_0=0$),  formula \eqref{ToHeatTr} gives an $\SL(2,\mathbb{R})$ group action on a solution $U(x,t)$ of the heat equation
\begin{equation}\label{SL2-inv-sol}
\begin{split}
   & u=C(c t+d)^{-1/2}(x-r)\exp\left[-\frac{d}{2\gamma^2(x-r)^2(c t+d)}\right]U(\tilde{x},\tilde{t}), \\
    & \tilde{x}=-\frac{\sqrt{2(a d-b c)}}{\gamma(c t+d)(x-r)}, \quad \tilde{t}=M(t)=\frac{a t+b}{c t+d}, \quad a d-b c> 0.
\end{split}
\end{equation}

Solution invariant under  the dilatational type vector field $v_2$ of \eqref{6-dim-c20}
$$v_2=t\gen t+\frac{1}{2}\sigma \psi\gen x+\frac{1}{4}(4c_0 t+\psi \sigma')u\gen u$$
assumes the form
\begin{equation}\label{ansatz}
  u(x,t)=R(\zeta)\sqrt{\sigma}e^{c_0 t},  \quad \zeta=\frac{\psi}{\sqrt{t}}.
\end{equation}
Substitution into \eqref{quadra-volat-2} and taking into account \eqref{2ndODE-2} we find
that $R$ should satisfy $R''+\zeta R'=0$ with solution
$$R=C_1 \erf\left(\frac{\zeta}{\sqrt{2}}\right)+C_2,$$
where $\erf(z)$ is the error function defined by
$$\erf(z)=\frac{2}{\sqrt{\pi}}\int_0^{z}e^{-\eta^2}d\eta.$$
Solution \eqref{ansatz} includes \eqref{inv-sol-2} for $C_1=0$. Taking $C_2=0$ we obtain the solution
\begin{equation}\label{inv-sol-3}
  u(x,t)=C \erf\left(\frac{\psi}{\sqrt{2t}}\right)\sqrt{\sigma}e^{c_0t}
\end{equation}
with the initial condition $u(x,0^+)=C\sqrt{\sigma}$.
Obviously, the error function solution of the heat equation is included in \eqref{inv-sol-3}.

Invariance under the projective vector field
$$v_3=t^2\gen t+t\sigma \psi\gen x+\frac{1}{2}[t(2c_0t-1)-\psi^2+t\psi \sigma']u\gen u$$
leads to the invariant solution
$$u=R(\zeta)\sqrt{\frac{\sigma}{\psi}}\exp\left[c_0 t-\frac{\psi^2}{2t}\right], \quad \zeta=\frac{\psi}{t}.$$
Substituting this into \eqref{quadra-volat-2} and using the relation $8c_0+\sigma'^2-2\sigma \sigma''=0$ we find that $R$ satisfies the second order linear ODE
$$4\zeta^2R''-4\zeta R'+3R=0$$ with general solution $R=C_1\zeta^{1/2}+C_2\zeta^{3/2}$. Finally, the corresponding solution is given by
\begin{equation}\label{inv-sol-v3}
  u=\curl{C_1 t^{-1/2}+C_2 t^{-3/2}\psi}\sqrt{\sigma}\exp\left[c_0 t-\frac{\psi^2}{2t}\right].
\end{equation}
  Obviously, the solution \eqref{inv-sol-v4} is included in this solution.

\end{example}

\begin{example}\label{CIR-ex}
We consider the Cox--Ingersoll--Ross (CIR) PDE
\begin{equation}\label{CIR}
  u_t=\sigma x u_{xx}+(m+n x)u_x, \quad \sigma, m, n>0, \quad x\in \mathbb{R^{+}}.
\end{equation}
We have $a=\sigma x$, $b=m+nx$, $c=0$. We start calculating the following quantities
$$I=2\sqrt{\frac{x}{\sigma}},  \quad J=\frac{1}{\sqrt{\sigma x}}\left[\frac{\sigma}{2}-(m+nx)\right],  \quad K=-\frac{1}{2\sigma x}[\frac{1}{2}(m+n x)^2-\sigma m].$$
We establish the condition ensuring that a non-trivial symmetry algebra exists
\begin{equation}
  K(x)=\frac{\mu}{I^2}+c_2 I^2+c_1 I+c_0.
\end{equation}
After some manipulation, this implies that
$$c_1=0,  \quad c_2=-\frac{n^2}{16},  \quad c_0=-\frac{mn}{2\sigma},  \quad \mu=-\frac{\Phi}{4\sigma^2},$$ where $\Phi=(2m-\sigma)(2m-3\sigma)$.
If $\mu\ne 0$, then the corresponding non-trivial symmetry algebra becomes four-di\-men\-sion\-al. Otherwise, if we let $\mu=0$ ($\Phi\equiv 0$) then we find that we must have $m\in\curl{\sigma/2,(3\sigma)/2}$ for the presence of a six-dimensional symmetry algebra. In the latter case, the corresponding CIR equation is isomorphic to the heat equation $u_t=u_{xx}$ by the action of the transformation formula \eqref{ToHeatTr}. We can write such transformations for both  of the values $m$ in a unified manner  for the special choice $\omega=0$ using \eqref{ToHeatTr}:
\begin{equation}\label{tr-t-x}
  \tilde{t}=T(t)=\frac{\alpha \cosh \frac{n t}{2}+\beta \sinh \frac{n t}{2}}{\gamma \cosh \frac{n t}{2}+\delta \sinh \frac{n t}{2}},  \quad \tilde{x}=\sqrt{\frac{-2n\Delta }{\sigma}}\left(\gamma \cosh \frac{n t}{2}+\delta \sinh \frac{n t}{2}\right)^{-1}\sqrt{x},
\end{equation}
\begin{equation}\label{tr-u}
  u=C\left(\gamma \cosh \frac{n t}{2}+\delta \sinh \frac{n t}{2}\right)^{-1/2}g(x)\exp\left[-\frac{n}{2\sigma}\left(1+\tanh(\frac{nt}{2}+\varepsilon)\right)x+c_0t\right]\tilde{u},$$
where $C$ is a nonzero arbitrary constant, $\tanh\varepsilon=\delta/\gamma$,  $c_0\in\curl{-n/4,-3n/4}$ and
$$g(x)=\begin{cases}
  1, &  m=\frac{\sigma}{2}, \; \\
  \frac{1}{\sqrt{x}}, & m=\frac{3\sigma}{2}.
\end{cases}
\end{equation}

An interesting solution to \eqref{CIR} is obtained if we choose $\tilde{u}=1$. We first choose the pair $(m,c_0)=(\sigma/2,-n/4)$, $\alpha=\delta=0$, $\beta=\gamma=1$ ($\Delta=-1$) and obtain the solution
\begin{equation}\label{sol1}
  u=C(1+e^{nt})^{-1/2}\exp\left[-\frac{n}{\sigma}\frac{x}{(1+e^{-nt})}\right].
\end{equation}
For the other possibility $(m,c_0)=(3\sigma/2,-3n/4)$, the solution produced from mapping to $\tilde{u}=1$ is
\begin{equation}\label{sol2}
  u=C(e^{2nt}+e^{nt})^{-1/2}x^{-1/2}\exp\left[-\frac{n}{\sigma}\frac{x}{(1+e^{-nt})}\right].
\end{equation}

If we have $m\not\in\curl{\sigma/2,(3\sigma)/2}$, then  transformation \eqref{tr-t-x}  with $g(x)=x^{(\sigma-2m)/(4\sigma)}$ maps the equation to the second canonical form
$$\tilde{u}_{\tilde{t}}=\tilde{u}_{\tilde{x}\tilde{x}}+\frac{\mu}{\tilde{x}^2},  \quad \mu=-\frac{(2m-\sigma)(2m-3\sigma)}{4\sigma^2}.$$

We refer the interested reader to Example 5.3 of \cite{Guengoer2019} for a detailed study of the non-trivial Lie point symmetries, relevant transformations and  their use for construction of the fundamental solution $u=K(x,t;x_0)$ satisfying the boundary condition $K(x,0;x_0)=\delta(x-x_0)$ in  the special case $n=0$.
\end{example}

Finally, we construct a fundamental solution for a multidimensional PDE invariant under rotation.

\begin{example}
We wish to solve the Cauchy problem for  the PDE
  \begin{equation}\label{multi-pde}
    u_t=\Delta u+\left(\frac{\mu}{|x|^2}+\omega^2 |x|^2\right)u=0,  \quad x\in{\mathbb{R}^n},  \quad \mu>0
  \end{equation}
with the condition
$$u(x,0;y)=\delta(x-y),  \quad y\in \mathbb{R}^n,$$
where $\Delta$ is the  Laplacian operator in $\mathbb{R}^n$, $|x|$ defines the Euclidean length in $\mathbb{R}^n$ and $\delta(x-y)$ is the Dirac distribution with source at $y$. The case $n=1$ was already treated in \cite{Guengoer2021}.

Fundamental solution should be invariant under rotations so that we look for a radial solution as fundamental solution in the form $u(x,t)=u(|x|,t)=u(r,t)$, where $r=|x|=(x_1^2+\cdots+x_n^2)^{1/2}$ is the radial variable. We require to solve the one-dimensional diffusion PDE
\begin{equation}\label{1-dim-pde}
  u_t=u_{rr}+\frac{n-1}{r}u_r+\left(\frac{\mu}{r^2}+\omega^2 r^2\right)u=0
\end{equation}
with the condition
\begin{equation}\label{init-cond}
  u(r,0;\rho)=\delta(r-\rho), \quad \rho=|y|.
\end{equation}
We are dealing with a PDE of the form \eqref{mainLPE} with coefficients
$$a(r)=1, \quad b(r)=\frac{n-1}{r}, \quad c(r)=\frac{\mu}{r^2}+\omega^2 r^2.$$ We need the following quantities
$$I=r,\quad J=-b,  \quad IJ=1-n,$$ and
$$K=\left[\mu-\frac{(n-1)(n-3)}{4}\right]r^{-2}+\omega^2 r^2.$$
The last relation implies that provided $\mu\ne (n-1)(n-3)/4$, the symmetry algebra must be four-dimensional.

Comparing this with \eqref{K} we have
$$c_0=c_1=0, \quad c_2=\omega^2,$$
and from \eqref{4-dim-c2p}, we find the basis
\begin{equation}\label{4-dim-L}
\begin{split}
&v_1=\gen t,\\
&v_2=\cos (4\omega t)\gen t-2\omega \sin (4\omega t)r\gen r+[n\omega \sin (4\omega t)+2\omega^2 \cos (4\omega t)r^2] u\gen u,\\
&v_3=\sin (4\omega t)\gen t+2\omega \cos (4\omega t)r\gen r+[-n\omega \cos (4\omega t)+2\omega^2 \sin (4\omega t)r^2] u\gen u,\\
&v_4=u\gen u
\end{split}
\end{equation}
with nonzero commutation relations
\begin{equation}\label{o21}
  [v_1,v_2]=-4\omega v_3,  \quad [v_1,v_3]=4\omega v_2,  \quad [v_2,v_3]=4\omega v_1.
\end{equation}
The subalgebra spanned by $v_1,v_2,v_3$ is  an $\So(2,1)$ algebra, which is isomorphic to the $\Sl(2,\mathbb{R})$ algebra. This is seen by making the change of basis
$$v_1'=4\omega(v_1+v_2), \quad v_2'=4\omega v_3 \quad v_3'=4\omega(v_1-v_2)$$
The vector fields $v_1',v_2',v_3'$ thus form an  $\Sl(2,\mathbb{R})$ algebra with the commutation relations
$$[v_1',v_2']=v_1',  \quad [v_1',v_3']=2v_2',  \quad [v_2',v_3']=v_3'.$$
The symmetry algebra has the direct-sum structure
$$\lie=\Sl(2,\mathbb{R})\oplus\curl{u\gen u}.$$

When $\mu= (n-1)(n-3)/4$, the PDE \eqref{1-dim-pde}
has  six-dimensional symmetry algebra  isomorphic to the heat algebra. In this case,  two additional symmetries exist. They are vector fields $v_4, v_5$  in the notation of \eqref{6-dim-c2p} given by
\begin{equation}\label{additional-symm}
\begin{split}
&v_4=\cos(2\omega t)\gen r+[\omega \sin (2\omega t)r-\frac{(n-1)}{2r} \cos (2\omega t)] u\gen u,\\
&v_5=\sin (2\omega t)\gen r-[\omega \cos (2\omega t)r+\frac{(n-1)}{2r} \sin (2\omega t)] u\gen u.
\end{split}
\end{equation}

We now pass to find the group invariant solution  preserving the boundary conditions $t=0$, $r=\rho$ and \eqref{init-cond}. With these conditions  applied to the vector field
$$v=\sum_{i=1}^4 k_iv_i,$$
we get the conditions
\begin{equation}\label{conds-k}
  k_1+k_2=0,  \quad k_3=0,  \quad k_4-2k_1 \omega^2 \rho^2=0.
\end{equation}
This leads to the subalgebra generated by the vector field
\begin{equation}\label{gen-vf-1}
  v=2\sin^2 (2\omega t)\gen t+2\omega \sin (4\omega t)r\gen r+\omega[2\omega \rho^2-2\omega \cos (4\omega t) r^2-n\sin (4\omega t)]u\gen u.
\end{equation}
We find  two functionally independent invariants of $v$ by solving the linear PDE $v(F)=0$ by the method of characteristics  as
\begin{equation}\label{inv-1}
  \eta=\frac{r}{\sin (2\omega t)},  \quad \zeta=u^{-1}{\sin (2\omega t)}^{-n/2}\exp\left[-\frac{\omega(r^2+\rho^2)}{2\tan(2\omega t)}\right].
\end{equation}
We make the ansatz
\begin{equation}\label{grp-inv-sol}
  u(r,t)=F(\eta){\sin (2\omega t)}^{-n/2}\exp\left[-\frac{\omega(r^2+\rho^2)}{2\tan(2\omega t)}\right].
\end{equation}
If $u$ solves the original PDE, $F$  satisfies an ODE of Bessel type
\begin{equation}\label{Bessel}
  \eta^2 F''+(n-1)\eta F'+(\mu-\omega^2\rho^2\eta^2)F=0.
\end{equation}
The change of variable $F=\eta^{(2-n)/2}G(\eta)$ transforms it to the modified Bessel equation
\begin{equation}\label{modified-Bessel}
  \eta^2 G''+\eta G'-(\eta^2+\nu^2)G=0,  \quad \nu^2=\frac{(n-2)^2}{4}-\mu
\end{equation}
with the solution
$$F(\eta)=\eta^{(2-n)/2}\left(c_1(\rho)I_{\nu}(\omega r \rho)+c_2(\rho)I_{-\nu}(\omega r \rho)\right),  $$
where $I_{\nu}(z)$ is the modified  Bessel function of the first kind of  order $\nu$ for non-integer $\nu$. Since $I_{-\nu}(z)$  is not integrable near zero for $\nu\geq 1$ then we will have to impose $c_2(\rho)=0$. With the choice $c_1(\rho)=\omega\rho^{(2-n)/2}$ we obtain the fundamental solution
\begin{equation}\label{fund-sol}
  u(r,t;\rho)=\frac{\omega(r\rho)^{(2-n)/2}}{\sin (2\omega t)}\exp\left[-\frac{\omega(r^2+\rho^2)}{2\tan(2\omega t)}\right]I_{\nu}\left(\frac{\omega r \rho}{\sin 2\omega t}\right).
\end{equation}

We note that when $\omega=0$ ($c_2=0$), formula \eqref{4-dim-c20} gives the symmetry algebra
\begin{equation}\label{omega0}
\begin{split}
&v_1=\gen t,\\
&v_2=t\gen t+\frac{r}{2}\gen r+\frac{(1-n)}{4} u\gen u,\\
&v_3=t^2\gen t+tr\gen r-\frac{1}{4}(r^2+2nt) u\gen u,\\
&v_4=u\gen u.
\end{split}
\end{equation}
The vector fields $v_1,v_2-v_4/4,v_3$ span an algebra isomorphic to $\Sl(2,\mathbb{R})$ algebra. The corresponding fundamental solution for the one-dimensional version of \eqref{1-dim-pde} can be shown to be (see \cite{Guengoer2018a} for a derivation)
\begin{equation}\label{fundsol-can}
  u(x,t;y)=\frac{\sqrt{xy}}{2t}\exp\curl{-\frac{x^2+y^2}{4t}}I_{\nu}
\left(\frac{xy}{2t}\right), \quad x>0,\; y>0,
\end{equation}
where $I_{\nu}$ denotes the modified Bessel function of  the first kind of order $\nu=\sqrt{1/4-\mu}$.

We note that the solution \eqref{fundsol-can} can be recovered from \eqref{fund-sol} by taking its limit as $\omega\to 0$.
\end{example}

\bigskip

\noindent{\bf Conflict of interest statement}
\smallskip

\noindent The author declares that there is no conflict of interest.

\medskip

\noindent{\bf Data availability statement}
\smallskip

\noindent The data that support the findings of this work are available within the article.


\end{document}